\documentstyle[aps,pra,epsf,multicol,array]{revtex}
\newcommand{\dket}[1]{ \left| #1 \right\rangle }

\newcommand{\ba}[1]{\begin{array}{#1} }
\newcommand{\ea}{\end{array}}
\newcommand{\rt}{ \frac{1}{\sqrt{2}} }
\newcommand{\bm}{\bf}

\begin{document}
\title{Mediated Entanglement And Correlations In A Star Network Of Interacting Spins}
\author{A. Hutton and S. Bose}
\address{Centre for Quantum Computation, Clarendon Laboratory,
    University of Oxford,
    Parks Road,
    Oxford OX1 3PU, England}
\maketitle

\begin{abstract}
We investigate analytically a star network of spins, in which all
spins interact exclusively with a central spin through Heisenberg
XX couplings of equal strength. We find that the central spin
correlates and entangles the other spins at zero temperature to a
degree that depends on the total number of spins. Surprisingly,
the entanglement depends on the evenness or oddness of this
number and some correlations are substantial even for an infinite
collection of spins. We show how symmetric multi-party states for
optimal sharing and splitting of entanglement can be obtained in
this system using a magnetic field.
\end{abstract}

\begin{figure}
\begin{center}
\leavevmode \epsfxsize=7cm \epsfbox{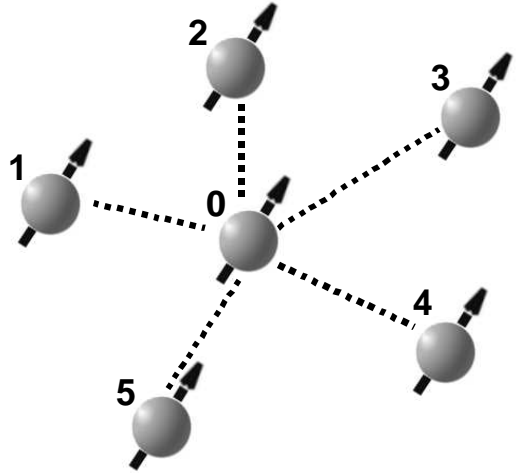}
 \caption{\narrowtext This figure depicts the star configuration of spins. The
spin labelled 0 is the central spin, which interacts by the XX
Heisenberg interaction with spins 1 - 5 around it.}
\label{piccy}
\end{center}
\end{figure}

\begin{multicols}{2}
For a long time spin correlations in 1D chains and higher
dimensional lattices of interacting spins have been a subject of
extensive interest \cite{bethe,kauf,lieb}. Recently, the same
systems have been studied from the point of view of truly {\em
quantum correlations} or entanglement
\cite{wot-ocon,thermentA,thermentB,wang01,gunlycke01,kamta,falci,wangxx,ibose}.
However, lattices of various dimensions are not the {\em only}
physical systems whose fabrication is possible with current
technology. In particular, with the advent of quantum computation,
various technologies have evolved which can make any member of an
array of qubits (systems isomorphic to spin-1/2) controllably
interact with any other member \cite{micro-traps,dots,cavity}. It
thus becomes possible to visualize structures of interacting
spins which do not fall into the category of lattices in various
dimensions. One very simple structure that one can imagine, is a
{\em spin-star}, as opposed to the extensively studied {\em
spin-chains}. In such a spin star, there is a preferred spin,
which we call the {\em central spin} which interacts with {\em
all} the other spins. All the non-central spins (which we will
call the {\em outer spins}), on the other hand, do not directly
interact among themselves. The structure is clearly depicted in
Fig.\ref{piccy}. $0$ depicts the central spin. The spins $1-5$
interact only with the central spin and not with each other. The
architecture is analogous to the star distribution networks used
in communications. To our knowledge, not just entanglement and
correlations, but also the statistical mechanics of such a
structure remains unexplored.

       The star configuration with couplings of equal strength has many
symmetry properties due to its invariance under the exchange of
any two outer spins and we solve it exactly in the case of an
Heisenberg XX interaction. The XX model was intensively
investigated for spin chains
 by Lieb, Schultz and Mattis \cite{lieb}, has been realized in recent years as an effective Hamiltonian in
some systems \cite {dots,cavity}. We find that the central spin
{\em mediates} correlations and entanglement between the outer
spins at zero temperature to a degree that depends on the total
number of spins in the spin-star. As expected, the entanglement
goes down on average with the increase of the total number of
spins. But it shows curious oscillations with the evenness or
oddness of this number. This is surprising, because in a star
network, addition of an extra outer spin in the network is only
expected to make it harder for the central spin to mediate
entanglement. For the same reason, it is also surprising that we
find that {\em irrespective} of the number of the outer spins,
the solitary central spin is capable of imposing substantial spin
ordering (correlation function $\geq 1/2$) in the X and Y
directions. We also show that we can apply a magnetic field to
our system to obtain multi-party states for optimal symmetric
splitting \cite{dur} and optimal symmetric sharing \cite{koashi}
of entanglement as the ground state and as a simple derivative of
the ground state respectively.

 The
Hamiltonian which describes our system is given by
\begin{equation}
H={\cal J} \left( \sigma_{0x}\sum_{\mbox{outer}} \sigma_{ix}  +
\sigma_{0y} \sum_{\mbox{outer}} \sigma_{iy} \right)
\label{eqn:firstH}
\end{equation}
where the summation over 'outer' refers to the outer spins,
$\sigma_{ix}$ and $\sigma_{iy}$ denote the $\sigma_x$ and
$\sigma_y$ Pauli operators for the $i$th outer spin and
$\sigma_{0x}$ and $\sigma_{0y}$ denote the $\sigma_x$ and
$\sigma_y$ Pauli operators for the central spin. It can be shown
that $J_x = (1/2)\sum_{\mbox{outer}} \sigma_{ix} $,$J_y =
(1/2)\sum_{\mbox{outer}} \sigma_{iy}$ and $J_z =
(1/2)\sum_{\mbox{ring}} \sigma_{iz}$ obey the standard angular
momentum commutation relations (we have taken $\hbar=1$). This
implies that the outer spins {\em collectively behave} as a
single spin with spin operator ${\bf J}=\hat{i} J_x+\hat{j}
J_y+\hat{k} J_z$. It can be shown that $J^2 = J_x^2 + J_y^2 +
J_z^2$ commutes with $H$. It will also help to define the total
angular momentum operator $\bm{F} = (1/2)\bm{\sigma_0} + \bm{J}$,
where $\bm{\sigma_0}=\hat{i} \sigma_{0x}+\hat{j}
\sigma_{0y}+\hat{k} \sigma_{0z}$ and it can be shown that the
z-component, $F_z$ obeys $[H, F_z] = 0 ,\: [J^2, F_z] = 0$.
Therefore simultaneous eigenstates of $H$, $J^2$ and $F_z$ can be
constructed.

 It is convenient to recast the Hamiltonian in Eq.\ref{eqn:firstH}
using the raising and lowering operators $\sigma_{\pm}  =
(\sigma_x \pm i\sigma_y)$ and \(J_{\pm} = (1/2)\sum_{\mbox{outer}}
\sigma_{\pm} \) as
\begin{equation}
H = {\cal J} \left( \sigma_{0+}J_- + \sigma_{0-}J_+ \right)
\label{eqn:Hladder}
\end{equation}
The above Hamiltonian thus represents a resonant interaction
between a spin-$(1/2)$ and a higher spin (with operator ${\bf J}$)
system. Inspired by the above form of $H$ (in particular, its
similarities with the Jaynes-Cummings model of quantum optics
\cite{knight}), we conjecture the following state as the general
form of its eigenstates
\begin{equation}
\rt \left( \dket{0}\dket{j,m} \pm \dket{1}\dket{j,m-1} \right)
\label{eqn:eigstate}
\end{equation}
where the first ket in each term denotes the central spin
($\dket{0}$ and $\dket{1}$ stand for the $\dket{-1/2}$ and
$\dket{1/2}$ spin states of the central spin), and the second ket
is an eigenstate of $J^2$. $j$ is the quantum number associated
with eigenstates of $J^2$ (eigenvalue of $J^2$ is $j(j+1)$), and
$m$ is the quantum number for $J_z$. The way $\dket{j,m}$ is
paired with central spin state $\dket{0}$ and $\dket{j,m-1}$ with
$\dket{1}$ in Eq.(\ref{eqn:eigstate}) means that the state is an
eigenstate of $F_z$ with eigenvalue $m-1/2$.
Eq.\ref{eqn:eigstate} is valid for $m = j$ to $m = -j+1$. There
are also two additional states where only one of the terms exist:
$\dket{1}\dket{j,j}$, because $\dket{0}\dket{j,j+1}$ does not
exist, and similarly $\dket{0}\dket{j,-j}$.

 We now seek to prove that states of the form in Eq.(\ref{eqn:eigstate}) are indeed eigenstates
by applying the Hamiltonian in Eq.\ref{eqn:Hladder} to them.
Applying $H$ gives
\begin{eqnarray*}
& & {\cal J}(1/\sqrt{2}) \left( \dket{1}J_-\dket{j,m} \pm \dket{0}J_+\dket{j,m-1} \right) \\
= & & {\cal J} (1/\sqrt{2}) \left( \dket{1}\sqrt{(j+m)(j-m+1)}\dket{j,m-1} \right. \\
& & \left. \pm \dket{0}\sqrt{(j-(m-1))(j+(m-1)+1)}\dket{j,m} \right) \\
= & & \pm {\cal J} \sqrt{(j+m)(j-m+1)} \\
& & \times (1/\sqrt{2}) \left( \dket{0}\dket{j,m} \pm \dket{1}\dket{j,m-1}  \right), \\
\end{eqnarray*}
where the standard relations for $J_{\pm}\dket{j,m}$ have been
used in the second step. This confirms that states in
Eq.\ref{eqn:eigstate} are eigenstates, with energy eigenvalues
\begin{equation}
E = \pm {\cal J} \sqrt{(j+m)(j-m+1)}
\label{eqn:energies}
\end{equation}
Next we prove that these states account for {\em all} possible
eigenstates, so that we have found a {\em complete eigen-basis}
for $H$.


 The total number of eigenstates of the form Eq.\ref{eqn:eigstate}
is $4j$ for a given value of $j$ (as $m$ runs from $j$ to $-j+1$
and there are two states for each $m$ due to the $\pm$ in
Eq.\ref{eqn:eigstate}). In addition, there are the two states
where one of the terms in Eq.\ref{eqn:eigstate} was missing.  In
total therefore, for a given value of $j$, there are $4j+2$
possible eigenstates. If we can now enumerate the possible values
of $j$ (which is the label for the total angular momentum of the
outer spins), we can find out the number of eigenstates we have
been able to account for.

 Suppose there are $N$ outer spins. Let us first consider the
 different ways of generating different $m$ values. Then there are $^NC_0=1$ ways of
having all these spins aligned in the same direction, giving $m =
N/2$.  If all the outer spins but one are aligned, then $m =
(N-1)/2 - 1/2$, and there will be $^NC_1$ ways of getting this
value of $m$.  For the general case of $N-r$ spins aligned and
$r$ spins anti-aligned to them, then $m = (N-r)/2 - r/2$ and
there are $^NC_r$ ways of getting this value of $m$.

 We now have to group the above $m$ values under suitable $j$ values.
Consider the case where $m = N/2$.  As there is only one
arrangement of spins giving rise to this, there should be only one
value of $j =  N/2$ to account for this. In other words, if there
was more than one allowed $j=N/2$ value for the collection of
spins, there would be more than one $m=N/2$, which we know, is not
the case.  Next consider $m = (N-2).\frac{1}{2}$. There are
$^NC_1$ such ways of obtaining this value of $m$.  One of these
ways will arise from when $j = N.\frac{1}{2}$, which means there
must be $^NC_1 - ^NC_0$ of $j = (N-2).\frac{1}{2}$ to account for
the remaining ways of getting this value of $m$. This procedure
for determining possible $j$ continues until all $m$ are
accounted for. In general, there are $^NC_r - ^NC_{r-1}$ ways of
obtaining $j = (N-2r).\frac{1}{2}$, with allowed values of r
ranging from $r = 0$ to $ r = N/2$ if $N$ is even, or $r =
(N-1)/2$ if $N$ is odd.

   Given these values and $j$, and that each value of $j$ can produce $4j+2$ eigenstates,
then a summation expressing the total number of states we have accounted for is
\begin{equation} \sum_{r=0}^x \left( ^NC_r - ^NC_{r-1} \right) \times
\left(4\left[(N-2r).\frac{1}{2} \right] + 2 \right) \end{equation}
where $x = N/2$ if $N$ is even, or $(N-1)/2$ if $N$ is odd.  The
result of the summation is determined by observing that
successive terms in summation cancel out some of the preceding
terms.  The result is independent of whether $N$ is even or odd,
and is $2^{N+1}$.  This means the total number of eigenstates of
the type we identified in Eq.\ref{eqn:eigstate} is $2^{N+1}$,
which is equal to the dimension of the Hilbert space of all the
$N$ outer spins plus the central spin. Therefore the states in
Eq.\ref{eqn:eigstate} (when combined with states of the type
$\dket{1}\dket{j,j}$ and $\dket{0}\dket{j,-j}$) describe
\emph{all} the eigenstates of $H$.


The ground state is the state with the lowest energy. To identify
it we will first have to assume a sign of ${\cal J}$, which we
take to be positive. All our results about entanglement and
correlations will be exactly the same for negative ${\cal J}$. In
Eq.\ref{eqn:energies} the lowest energy is obtained when $j$ has
its maximum possible value and $m$ has its minimum possible
value. For the case $N$ odd, the lowest energy is when
$m=\frac{1}{2}$ i.e. the eigenstate
\begin{equation} (1/\sqrt{2}) \left( \dket{0}\dket{N/2,1/2} -
\dket{1}\dket{N/2,-1/2} \right)
\end{equation}
and if $N$ is even, then in fact the ground state is degenerate
because there are two states with the lowest possible energy,
when $m=0$ or $m=1$. These are
\begin{eqnarray}
\rt \left( \dket{0}\dket{N/2,0} - \dket{1}\dket{N/2,-1} \right) \nonumber\\
\rt \left( \dket{0}\dket{N/2,1} - \dket{1}\dket{N/2,0} \right)
\end{eqnarray}
The reason for the above difference between $N$ even and $N$ odd
is that the two cases lead to an integral and half integral value
of $j$ respectively. When $j$ is half integral, $m=\pm 1/2$ is
allowed and gives an unique ground state. For $j$ integral, the
$0,-1$ and $1,0$ form two distinct $j,m$ pairs to combine with the
central spin-$1/2$ particle to give two degenerate ground states.

  To compute entanglement and correlations, it is useful to have
expressions in terms of the states of the individual outer spins
for these ground states. Let $\dket{0}$ and $\dket{1}$ stand for
the $\dket{-1/2}$ and $\dket{1/2}$ spin states of any outer spin.
For $N$ odd, the state $|N/2,1/2\rangle$ is an equal
superposition of all states with $(N+1)/2$ ones and $(N-1)/2$
zeros with no relative phase between them. The state
$|N/2,-1/2\rangle$ is the same type of state with $(N-1)/2$ ones
and $(N+1)/2$ zeros. For example, for $N=3$,
\begin{eqnarray}
 |3/2,1/2\rangle&=\frac{1}{\sqrt{3}} \left[ \dket{011} + \dket{101} + \dket{110} \right]  \nonumber\\
 |3/2,-1/2\rangle&=\frac{1}{\sqrt{3}} \left[ \dket{100} + \dket{010} + \dket{001}
\right]
\end{eqnarray}
There are similar expressions for the ground state for $N$ even.
The $\dket{N/2,0}$ state is an equal superposition of all states
with an equal number of zeros and ones, with no relative phase
between the superposed states. $\dket{N/2,\pm 1}$ is the same type
of state with $N/2\pm 1$ ones and rest zeros.

   Given the ground state, we are able to calculate the
entanglement between any two outer spins in this state ({\em
i.e.} at zero temperature). The symmetry of the problem implies
that the entanglement will be the same between \emph{any} two
outer spins. Since $H$ and $F_z$ commute, the form of the reduced
density matrix for any two spins is \cite{wot-ocon}
\[ \left( \ba{cccc}
v & 0 & 0 & 0 \\
0 & w & z & 0 \\
0 & \bar{z} & x & 0 \\
0 & 0 & 0 & y \\
\ea \right) \]
in the standard basis $ \{ \dket{00}, \dket{01}, \dket{10},
\dket{11} \}$.  For such density matrices, a measure of
entanglement called the concurrence \cite{entform} is given by
\cite{wot-ocon}
\[ C = 2\max\{|z| - \sqrt{vy},0\} \]
For $N$ odd, the concurrence comes out as
\[ C = 2\max\{\frac{1}{2N},0\} = \frac{1}{N} \]
For the case of $N$ even, where there are two ground states, a
similar procedure is followed, except that the reduced density
matrix is now described as an equal mixture of the two states.
This gives the concurrence as
\[ C = 2\max\{\frac{1}{2N} - \frac{1}{2N(N-1)} ,0\} = \frac{1}{N} - \frac{1}{N(N-1)} \]
Thus the entanglement goes to zero as $N\rightarrow \infty$,
which is expected, as the entanglement is {\em mediated} by the
central spin. The total entangling capability (and thereby
mediating capability) of the central spin is divided among a
larger and larger number of outer spins as $N$ becomes larger.
However, on going from an even $N$ to an odd $N+1$ number of
outer spins, the concurrence rises from $1/N-1/(N^2-N)$ to
$1/(N+1)$. Thus on increasing $N$ there are {\em curious
oscillations in concurrence} of amplitude $2/\{N(N-1)(N+1)\}$.
These oscillations, due to the different expressions for
concurrence in the cases of even and odd $N$, is due to {\em
mixedness} of the zero temperature density matrix for even $N$.
On application of a magnetic field in the $+Z$ direction, the
state $\left( \dket{0}\dket{N/2,0} - \dket{1}\dket{N/2,-1}
\right)$ becomes the {\em non-degenerate} ground state for even
$N$ and the oscillations in entanglement disappear.

There are also interesting results to be noted for correlation
functions in the star network. The $\langle \sigma_{1z}\sigma_{2z}
\rangle$ correlations follow the same pattern as the
entanglement, but
\begin{eqnarray*}
\left< \sigma_{1x}\sigma_{2x} \right> &=& \frac{1}{2} + \frac{1}{2N} \mbox{~for odd $N$}\\
\left< \sigma_{1x}\sigma_{2x} \right> &=& \frac{1}{2} +
\frac{1}{2N} - \frac{1}{2N(N-1)} \mbox{~for even $N$}.
\end{eqnarray*}
What is particularly surprising above, is the non-vanishing nature
of the correlations in the {\em large} $N$ limit. The solitary
central spin is capable to imposing spin order in the $X$
direction (so that $\left< \sigma_{1x}\sigma_{2x} \right> =
1/2$), even when there are an infinite number of outer spins to
order. The same result holds for $\left< \sigma_{1y}\sigma_{2y}
\right>$. Thus our system provides an effective way of imposing
order {\em simultaneously} in the $X$ and $Y$ directions for an
infinite collection of spins. This result also highlights a
crucial difference between entanglement and correlations: while a
finite dimensional quantum system cannot be individually
entangled to each member of an infinite collection of systems, it
can indeed be correlated individually to each of them.


We now show that the application of a magnetic field allows us to
change the ground state to
\begin{equation} \ba{lll}
\dket{\alpha} &=& \rt \left( \dket{0}\dket{N/2, -N/2+1} -
\dket{1}\dket{N/2,-N/2} \right)\nonumber \\
&=& \rt \left( \dket{0}\{\dket{000 \ldots 1}\} - \dket{1}\dket{000
\ldots 0} \right)
  \ea \label{eqn:nstate}
\end{equation}
where $\{\dket{000 \ldots 1}\}$ is a normalized state that is an
equal superposition of all states with  only one $\dket{1}$ with
no relative phase between the superposed components. This state
has the special significance that the concurrence between the
central spin and each of the outer spins is $1/\sqrt{N}$, which
is the {\em maximum} consistent with symmetric splitting of the
total entanglement of one qubit with a collection of $N$ qubits
among the $N$ qubits \cite{dur}. To our knowledge, this is the
first identification of the canonical state $|\alpha\rangle$ as
the ground state of an interacting spin system. Once the ground
state $|\alpha\rangle$ is generated, if the central spin is
measured and found to be in the state $\dket{0}$, the rest of the
spins are projected onto the state $\{\dket{000 \ldots 1}\}$. The
central spin can now be removed to make the state dynamically
steady (except for decoherence and spontaneous decay effects).
This state has the property that the concurrence between any two
spins is $2/N$, which is the {\em maximum} possible entanglement
in a collection of $N$ spins in which all pairs of spins are
equally entangled \cite{koashi}. While in Ref.\cite{thermentB},
it was only conjectured that a state of the type $\{\dket{000
\ldots 1}\}$ could be made the ground state of the isotropic
closed Heisenberg chain using a magnetic field, here we will
rigorously prove the preparation of $|\alpha\rangle$ and thereby
$\{\dket{000 \ldots 1}\}$.

 To show that it is possible to make $\dket{\alpha}$ the ground
state, we consider the energy eigenvalues in a uniform magnetic
field $B$ in the $+Z$ direction (in which the eigenstates remain
unchanged)
\begin{equation}
E = \pm {\cal J} \sqrt{(j+m)(j-m+1)} + (m-\frac{1}{2})B
\label{eqn:energyB}
\end{equation}
When $B$ becomes so high that the second term in $E$ dominates,
the relative ordering of the energy levels will be determined
purely by $m$, with the ground state being state with
$m=\frac{-N}{2}$, {\em i.e.} $\dket{\beta} = \dket{0}\dket{000
\ldots 0}$. It is straightforward to show that $\dket{\beta}$ has
an energy lower than $\dket{\alpha}$ for $B>{\cal J}\sqrt{N}$.
Before $\dket{\beta}$ becomes the ground state, there is a range
of $B$ in which $\dket{\alpha}$ is the ground state. To prove
this, we will have to show that the energy of $\dket{\alpha}$
will be less than that of all other states in a certain range.
The value of $B$ for which $\dket{\alpha}$ has the same energy as
a general state described by $j$ and $m$ is given by
%
\begin{equation}
B = \frac{ {\cal J} \left( \sqrt{N} - \sqrt{(j+m)(j-m+1)} \right)
}{ -(\frac{N}{2} - 1) - m} \label{eqn:equal2}
\end{equation}
$B$ attains its largest value, when $j$ is a maximum and $m$ has
it's most negative value. This happens when $j = \frac{N}{2} $
and $m = -\frac{N}{2} + 2$ for which $B = {\cal J} \sqrt{N}
\left(  \sqrt{2(1-1/N)} - 1 \right) < {\cal J} \sqrt{N}$.
%
Therefore, for a magnetic field with a range of $ {\cal J}
\sqrt{N} \left( \sqrt{2(1-1/N)} - 1 \right) < B < {\cal J}
\sqrt{N}$, the state $\dket{\alpha}$ is the ground state.

      In this letter we have introduced and solved a spin-star,
an architecture of interacting spins which cannot be classified
as a lattice in any dimension. It is physically realizable in
various arrays of qubits designed for quantum computation.
Testing for the entanglement and correlations predicted here
would serve as a benchmark test for the functioning of arrays of
qubits. If spin-spin interactions can be extended to long
distances, the spin-star could be used for distribution of
entanglement through the multiparty states for optimal symmetric
sharing and splitting of entanglement. Exploration of the full
statistical mechanics of a spin-star, would be interesting future
work.

AH thanks the UK EPSRC (Engineering and Physical Sciences
Research Council) for financial support.

\end{multicols}

\end{document}